\documentstyle[psfig,12pt]{article}
\newcommand{\HI}{H\thinspace\protect\footnotesize I\protect\normalsize}

\newcommand{\tfr}{Tully\,--\,Fisher relation}

\newcommand{\kms}{\,km\,s$^{-1}$}
\newcommand{\etal}{{\it et~al.}}
\newcommand{\cf}{{\it cf.\,}}
\newcommand{\eg}{{\it e.g.},\ }         
\newcommand{\ie}{{\it i.e.},\ }         

\def\la{\mathrel{\hbox{\rlap{\hbox{\lower4pt\hbox{$\sim$}}}\hbox{$<$}}}}
\def\la{\mathrel{\hbox{\rlap{\hbox{\lower4pt\hbox{$\sim$}}}\hbox{$<$}}}}
\def\ga{\mathrel{\hbox{\rlap{\hbox{\lower4pt\hbox{$\sim$}}}\hbox{$>$}}}}

\def\deg{{^\circ}}
\def\arcmin{\hbox{$^\prime$}}

\def\fm{\hbox{$.\!\!^{\rm m}$}}

\def\fdg{\hbox{$.\!\!^\circ$}}
\def\farcm{\hbox{$.\mkern-4mu^\prime$}}

\def\reference{\parskip 0pt\par\noindent\hangindent 0.5 truecm}

\begin{document}

\title{An Overview of Uncovered and Suspected Large-Scale Structures behind the Milky Way}

\author{Ren\'ee C. Kraan-Korteweg $^{1}$ \and
 Patrick A. Woudt $^{2}$
} 

\date{}
\maketitle

{\center
$^1$ Depto. de Astronom\1a, Univ. de Guanajuato, Apartado
Postal 144, Guanajuato GTO 36000, Mexico\\kraan@astro.ugto.mx\\[3mm]
$^2$ ESO, Karl-Schwarzschildstr. 2, D-85748 Garching bei M\"unchen,
Germany\\pwoudt@eso.org\\[3mm]
}

\begin{abstract}
Various dynamically important extragalactic large-scale structures 
in the local Universe lie behind the Milky Way.
Most of these structures (predicted and unexpected) have only 
recently been made ``visible'' through dedicated deep
surveys at various wavelengths. The wide range of 
observational searches (optical, near infrared, far infrared, radio and X-ray)
for galaxies in the Zone of Avoidance (ZOA) 
will be reviewed and the uncovered 
and suspected large-scale structures summarised. Particular emphasis 
is given to the Great Attractor region where the existence of yet
another cluster is suspected (Woudt 1998). Predictions from
reconstructions of the density field in the ZOA
are discussed and compared with observational evidence. Although
no major structures are predicted out to about $v \la 10000$~\kms\
for which no observational evidence exists, the comparison between
reconstructed density fields and the observed galaxy distribution 
remain important as they allow derivations of the density and
biasing parameters.
\end{abstract}

 
\bigskip

\section{Introduction}
In the last few years, a lot of progress has been achieved in
uncovering the galaxy distribution behind the Milky Way through various
multi-wavelength approaches. These sometimes quite tedious tasks are
necessary in order to understand the
dynamics in the nearby Universe and answer the question
whether the perturbations on the smooth Hubble expansion 
such as the dipole in the Cosmic Microwave 
Background and other velocity flow fields can be fully explained
by the irregular galaxy distribution, respectively mass distribution
if galaxies are fair tracers of mass. 
In section 2, the various observational methods are described and
the large-scale structures (LSS) uncovered to date summarised.  
With respect to \HI\ and near infrared surveys, only their characteristics
-- advantages, limitations and selection effects -- in context 
to other approaches are discussed as specific surveys and results are 
presented in detail elsewhere in this volume (\cf\ Henning \etal, 
Rivers \etal, Juraszek \etal, and Schr\"oder \etal). 
In the last section, theoretical predictions applied to the recent 
deeper sampled galaxy surveys covering volumes out to
$v \le 10000$~\kms\ are reviewed, to see whether they give new
indications of unknown or unsuspected prominent structures in the ZOA.  

 
\section{Observational Surveys in the Zone of Avoidance}
\subsection{Optical Surveys}
Systematic optical galaxy catalogs are generally limited to the 
largest galaxies (typically with diameters 
D $\ga 1\arcmin$, \eg\ Lauberts 1982). These catalogs become,
however, increasingly incomplete as the dust thickens, 
creating a ``Zone of Avoidance'' in the distribution of galaxies
of roughly 25\% of the sky. Systematically deeper searches for partially 
obscured galaxies --  down to fainter magnitudes and smaller 
dimensions compared to existing catalogs -- were performed
with the aim of reducing this ZOA. These surveys are not biased 
with respect to any particular morphological type. 

Most of the ZOA has meanwhile been surveyed (cf.
Fig.~1, in Kraan-Korteweg 1998), revealing many galaxy overdensities
uncorrelated with the patchy, optical extinction distribution.
Analysing the galaxy density as a function of the galaxy size, 
magnitude and/or morphology in combination with the foreground 
extinction has led to the identification of various important 
large-scale structures and their approximate distances.
Although the ZOA has been considerably reduced in this way (to about
10\% of the sky),
this method does not find galaxies in the thickest extinction
layers of the Milky Way, \ie\ where the optical extinction exceeds 
4 -- 5 magnitudes, respectively at Galactic latitudes below $b \la \pm 5\deg$.

Redshift follow-ups of well-defined samples are important
in mapping the large-scale structures in redshift space. So far, this has
been pursued extensively in the Perseus-Pisces (PP) supercluster 
area and in large parts of the southern ZOA. 
The prominant new galaxy structures revealed in this way are
summarised below. Their approximate positions (ordered in Galactic 
latitude) are given as ($\ell, b, v$), with $v$ in units of \kms:

$\bullet$ Behind the Galactic Bulge at ($0\fdg5, 9\fdg5, 8500$),
 Wakamatsu \etal\ (1994) identified
the rich Ophiuchus cluster (or supercluster) with some evidence of
it being linked to the adjacent slightly more distant Hercules cluster.

$\bullet$ At ($\ell, b$)$\sim$ ($33\deg, 5\deg-15\deg$), Marzke \etal\ (1996)
and  Roman \etal\ (1998) found evidence for a nearby cluster
close to the Local Void at 1500~\kms, as well as 
a prominent cluster behind the Local Void at 7500~\kms. The nearby cluster
is independently supported by data from the blind ZOA \HI-survey 
(Henning \etal, this volume).

$\bullet$ The connection of the Perseus-Pisces supercluster across 
the ZOA to the cluster A569, suspected by Focardi \etal\ in 1984, was 
confirmed by Chamaraux \etal\ (1990) and Pantoja \etal\ (1997).
The Perseus-Pisces chain folds back into the ZOA at higher redshifts at
($195\deg, -10\deg, 7500$), Marzke \etal\ (1996), Pantoja \etal\ (1997).

$\bullet$ In 1992, Kraan-Korteweg \& Huchtmeier uncovered a nearby cluster
in Puppis ($245\deg, 0\deg, 1500$) which was later shown by Lahav \etal\
(1993) to contribute a not insignificant component to the peculiar 
z-motion of the Local Group.

$\bullet$ Kraan-Korteweg \etal\ (1994) presented evidence for a continous
filamentary structure extending over $30\deg$ on the sky from the
Hydra and Antlia clusters across the ZOA, intersecting the Galactic
Plane at ($280\deg, 0\deg, 3000$). At the same longitudes, they noted 
significant clustering at $\sim$ 15000~\kms, indicative of a connection 
between the Horologium and Shapley clusters a hundred degrees apart in the sky.

$\bullet$ Kraan-Korteweg \& Woudt (1993) uncovered a shallow but extended
supercluster in Vela at ($285\deg, 6\deg, 6000$).

$\bullet$ Next to the massive cluster A3627 at the core of the Great 
Attractor (clustering in the Great Attractor region is discussed in 
the next section), Woudt (1998) 
discovered a cluster at ($306\deg, 6\deg,6200$) called the Cen-Crux 
cluster, and a more distant cluster, the Ara cluster at ($329\deg, 
-9\deg, 15000$). The latter might be connected to the 
Triangulum-Australis cluster.

\subsubsection{Clustering within the Great Attractor Region?}

Recent consensus is that the Great Attractor (GA) is probably an 
extended region ($\sim 40\deg {\rm x} 
40\deg$) of moderately enhanced galaxy density (Lynden-Bell 
1991, Hudson 1994) centered behind the Galactic Plane
at ($\ell, b, v$) $\sim$ ($320\deg, 0\deg,
4500$) (Kolatt \etal\ 1995).

Based on a deep optical galaxy search and subsequent redshift 
follow-ups,  Kraan-Korteweg \etal\ (1996) and Woudt (1998)
have clearly shown that the Norma cluster, A3627, at 
($325\fdg3, -7\fdg2$, 4882) is the most massive galaxy cluster 
in the GA region known to date and probably marks
the previously unidentified but predicted density-peak 
at the bottom of the potential well of the GA overdensity.
The prominence of this cluster has independently been 
confirmed by ROSAT observations: the 
Norma cluster ranks as the 6$^{th}$ brightest X-ray cluster in 
the sky (B\"ohringer \etal\ 1996). It is comparable in 
size, richness and mass to the well-known Coma cluster.
Redshift-independent distance determinations (R$_{\rm C}$ 
and I$_{\rm C}$ band Tully-Fisher analysis) of 
the Norma cluster have shown it to be at rest with respect 
to the rest frame of the Cosmic Microwave Background (Woudt 1998).

One cannot, however, exclude the possibility that other unknown rich 
clusters reside in the GA region.
Finding a hitherto uncharted, rich cluster of galaxies at the 
heart of the GA would have serious implications for our current 
understanding of this massive overdensity in the local Universe.
Woudt (1998) found various indications that PKS1343-601, the
second brightest extragalactic radio source in the southern sky 
($f_{20cm} = 79$~Jy, McAdam 1991, and references therein) might 
form the center of yet another highly obscured rich cluster. 

At ($\ell, b) \sim (309\fdg7, 1\fdg7$), this radio galaxy lies behind an
obscuration layer of about 12 magnitudes of extinction in 
the B-band, as estimated from the DIRBE/IRAS extinction maps (Schlegel 
\etal\ 1998). Its observed diameter of 28 arcsec in the Gunn-z filter 
(West \& Tarenghi 1989) translates into an extinction-corrected
diameter of 232 arcsec (following Cameron 1990). With a recession 
velocity of v = 3872~\kms\ (West \& Tarenghi 1989) this galaxy can 
be identified with a giant elliptical galaxy.
 
PKS1343-601 has recently been observed in the X-ray band with the 
ASCA satellite (Tashiro \etal\ 1998). This source is not detected 
in the ROSAT All Sky  Survey due to the large foreground extinction, 
i.e.~the soft X-ray emission is totally absorbed. However,
extended diffuse hard X-ray emission at the position of PKS1343-601
has been detected with ASCA. The excess flux, kT = 3.9~keV, is far 
too large for it being associated with a galactic halo surrounding the 
host galaxy, hence it might be due to the Inverse Compton process
-- or indicative of emission from a cluster.
  
As this prospective cluster is so heavily obscured, little data are
available to substantiate the existence of this cluster.
In Figure~1, a comparison of the A3627 cluster at
($325\fdg3, -7\fdg2, 4882$) and a mean extinction
in the blue of $1\fm5$ is compared to the prospective PKS1343 cluster
at ($309\fdg7, +1\fdg7, 3872$) with an extinction of 12$^{\rm m}$.
The top panel shows both sky distributions. 
One can clearly see that at 
the low Galactic latitude of the suspected cluster PKS1343, the optical 
galaxy survey could not retrieve the underlying galaxy distribution, 
especially not within the Abell radius (the inner circle in the top
right panel of Figure 1) of the suspected cluster. If PKS1343-601 marks
the dynamical center of the cluster, then the Abell radius, defined as 
1$\farcm$7/$z$ where $z$ is the redshift, corresponds to
2.2$\deg$ on the sky at the redshift-distance of PKS1343-601.

Interestingly enough, the shallow blind
ZOA-Multibeam \HI\ survey (Henning \etal, this volume) picks up a number
of prospective cluster members even though the shallow survey is
sensitive only to the most \HI-rich galaxies at the cluster velocity: 
over 60\% of the galaxies in the shallow
survey with velocities from 3000 to 5000~\kms\ lie in the
cluster area, \ie within 13\% of the area covered by the shallow survey.

\begin{figure}[tb]
\begin{center}
\hfil \psfig{file=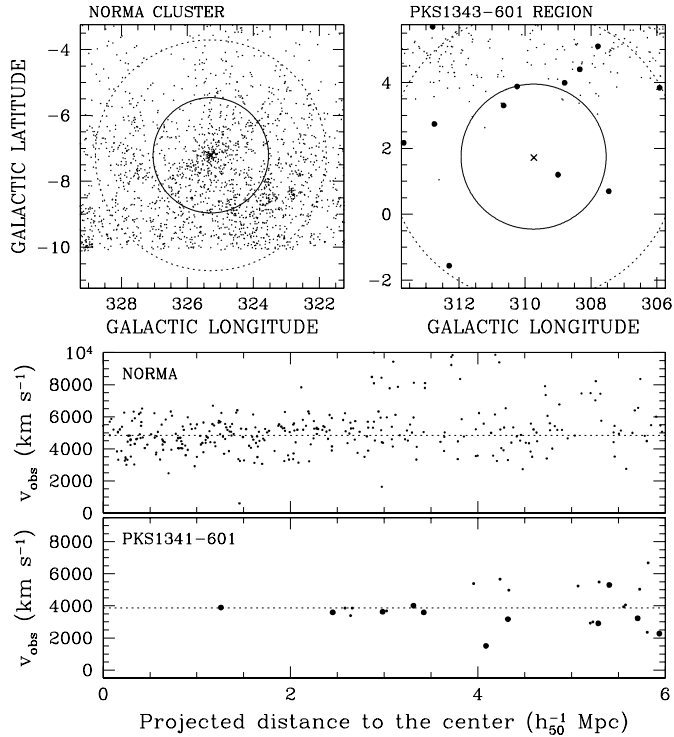,height=10cm} \hfil
\caption{A comparison of the rich A3627 cluster (A$_{\rm B}\sim
1\fm5$) and the suspected PKS1343 cluster (A$_{\rm B}\sim
12^{\rm m}$) in the GA region. Small dots are optically identified galaxies,
large dots galaxies detected in the shallow Multibeam ZOA survey.
The top panels display the sky distribution,where the inner
circle marks the Abell radius R$_{\rm A}$ = 3 h$_{50}^{-1}$
Mpc, the bottom 
panel the redshift distribution as a function of distance to 
the central radio source.
}
\label{pks1343}            
\end{center}
\end{figure}

The velocity 
distribution as a function of distance from the cluster center
for the PKS1343-601 region (bottom panel) provides further evidence 
for the existence of this cluster.
All measured velocities lie within a narrow range of the central
radio source, showing a similar distribution as in the Norma cluster.
One of the first data cubes from the full sensitivity
Multibeam ZOA survey that has been finished covers the 
prospective PKS1343 cluster area. A quick inspection gives 
further support for this prospective cluster: between
$3500 < v < 4000$~\kms\ a statistically
significant peak is evident in the velocity distribution 
(Juraszek \etal, 1998, priv. commun.).

We will image the prospective cluster within its Abell radius
in the near infrared (Woudt \etal\ in progress). These observations will 
allow us to determine whether or not PKS1343-601 is embedded 
in a centrally condensed overdensity of galaxies, comparable 
to the rich and massive Norma cluster.

\subsection{IRAS Surveys}
The IRAS Point-Source Catalog (PSC) has been exploited in the
last decade to identify galaxy candidates behind the ZOA. 
However, confusion with Galactic sources at low Galactic
latitudes still leave a considerable ``ZOA'' of over 10\%. 
Moreover, bright spiral and starburst galaxies dominate these 
samples, and cluster cores are hardly visible.

The advantage of using the IRAS survey for LSS studies are
the homogeneous sky coverage (all data from one instrument), and the 
various systematic redshift follow-ups, complete to given flux limits, 
\ie\ 2658 galaxies to f$_{60\mu m} = 1.9$~Jy (Strauss \etal\ 1992), 
     5321 galaxies to f$_{60\mu m} = 1.2$~Jy (Fisher \etal\ 1995), and 
$\sim$ 15000 galaxies to f$_{60\mu m} = 0.6$~Jy (Saunders \etal, in prep.). 

Considerable improvement towards filling the ZOA has been made 
through the confirmation of about 1000 IRAS galaxy candidates 
in the ZOA from K-band snapshots (Saunders \etal\, in prep.).

Using the IRAS survey, dedicated searches for large-scale
clustering within the whole ZOA ($|b| \le 15\deg$) have been made 
by Japanese groups (\cf\ Takata \etal\ 1996, for a summary). 
They used IRAS color criteria to select galaxy candidates which
were subsequently verified through visual examination on sky surveys 
such as the Palomar Observatory Sky Survey (POSS) of the northern
hemisphere and the ESO/SRC (United Kingdom Science Research Council)
Southern Sky Atlas. Because of their verification procedure, this 
data-set suffers the same limitations in highly obscured regions as optical
surveys. 

Based on redshift follow-ups of this ZOA IRAS galaxy sample, they
established various filamentary features and connections
across the ZOA. Most coincide with the structures described in
section 2.1. Both crossings of the Perseus-Pisces arms into the ZOA 
are very prominent -- considerably stronger in IRAS 
compared to optical data -- and the Puppis, Hydra, Centaurus and 
A3627 connections are clearly visible. They furthermore 
identified a new structure: the Cygnus-Lyra filament at 
($60\deg-90\deg, 0\deg, 4000$).

\subsection{HI Surveys}
In the regions of the highest obscuration and infrared confusion,
the Galaxy is fully transparent to the 21-cm line radiation 
of neutral hydrogen. \HI-rich galaxies can readily be found 
at lowest latitudes through the detection of their redshifted 
21-cm emission. Early-type galaxies generally are very gas-poor
and will not be uncovered in \HI\ surveys. Furthermore,
low-velocity extragalactic sources (blue- and red-shifted) 
within the strong Galactic \HI\ emission will
be missed, and -- because of baseline ripple -- galaxies 
close to radio continuum sources may also be missed.

As demonstrated by the first results from systematic 
\HI-surveys (\cf\ Henning \etal, Rivers \etal, and Juraszek \etal\
in this volume), these surveys clearly are very powerful in tracing 
spiral and \HI-rich dwarf galaxies through the deepest extinction
layer of the Milky Way. In particular, the results from the deep
Multibeam ZOA survey will be very exciting as they will trace the
galaxy distribution across the ZOA to a depth of $\sim$ 10000~\kms\
(\cf\ Fig.~2 in Kraan-Korteweg \etal, 1998).

\subsection{Near Infrared Surveys}
Near infrared (NIR) surveys are sensitive to early-type galaxies, tracers of 
massive groups and clusters missed in IRAS and \HI\ surveys, and have 
little confusion with Galactic objects. Moreover, they
are less affected by absorption than optical surveys.
Here, the recent NIR surveys such as 2MASS (Skrutskie \etal\ 1997) 
and DENIS (Epchtein 1997) provide a new tool to probe the ZOA.

First results from DENIS data are very promising (\cf\ Schr\"oder 
\etal, this volume). They are complementary to other surveys in the 
sense that they finally uncover
early-type galaxies at low Galactic latitudes ($|b| \ga 1 - 1\fdg5$).
Furthermore, a fair fraction ($\sim$~65\%) of the heavily obscured 
spiral galaxies 
detected in blind \HI\ surveys can be reidentified on DENIS images.
The combination of \HI\ data with NIR data allows the study of
the peculiar velocity field via the NIR \tfr\ ``in the ZOA'' 
compared to earlier interpolations of data ``adjacent to the ZOA'' and
this will, for instance, provide important new input for density field
reconstructions in the ZOA (\cf\ section~3).

\subsection{X-ray Surveys}
The Milky Way is transparent to the hard X-ray emission, \ie\ above
0.5-2.0 keV. Rich clusters generally are strong X-ray emitters.
Hence, the  X-ray surveys such as HEAO-1 and the ROSAT All Sky Survey
provide an optimal tool to search for clusters of galaxies at low 
Galactic latitude. So far, this possibility has not yet been 
pursued in any systematic way, even though a large number of 
X-ray bright clusters (\eg PKS0745-191) are located at low Galactic 
latitude (\cf\ Fabian 1994). This tool is of particular interest because 
it can unveil cores of clusters as, for instance, the suspected
cluster surrounding PKS1343-601. These are dominated by early-type 
galaxies and therefore difficult to trace in other wavelengths.

\section{Theoretical Reconstructions}
Various mathematical methods exist to reconstruct the galaxy 
distribution in the ZOA without having access to direct observations.

One possibility is the expansion of galaxy distributions
adjacent to the ZOA into spherical harmonics to recover
the structures in the ZOA, either with 2-dimensional catalogs (sky
positions)  or 3-dimensional data sets (redshift catalogs). 

A statistical method to reconstruct structures behind the Milky Way is
the Wiener Filter (WF), developed explicitly for reconstructions of
corrupt or incomplete data (\cf\ Lahav 1994, Hoffman 1994). Using the 
WF in combination with linear theory allows the determination
of the real-space density of galaxies, as well as their velocity
and potential fields.

The POTENT analysis developed by Bertschinger \& Dekel (1989)
can reconstruct the potential field (mass distribution) from
peculiar velocity fields in the ZOA (Kolatt \etal\ 1995). The 
reconstruction of the potential fields versus density fields 
have the advantage that they can locate hidden overdensities  (their 
signature) even if ``unseen''. 

Because of the sparsity of data and
the heavy smoothing applied in all these methods, only structures 
on large scales (superclusters) can be mapped.
Individual (massive) nearby galaxies that can perturb the
dynamics of the Universe quite locally (the vicinity of the Local
Group or its barycenter) will not be uncovered in this manner. 
But even if theoretical methods can outline LSS
accurately, the observational efforts do not become
superfluous. The comparison of the real galaxy distribution 
$\delta_{g}$ ({\bf r}), from \eg\ complete redshift surveys, with the
peculiar velocity field {\bf v}({\bf r}) will lead to an estimate of the 
density and biasing parameter ($\Omega^{0.6}/{b}$)  through the
equation 
\begin{equation}
 \nabla \cdot {\bf v (r)} = - {{\Omega^{0.6}}\over{b}} \, \, \delta_g
({\bf r}),
\end{equation}
\cf\ Strauss \& Willick (1995) for a detailed review.

\subsection{Early Predictions}
Early reconstructions on relatively sparse data galaxy catalogs have
been performed within volumes out to $v \le$ 5000~\kms. Despite
heavy smoothing, they have been quite successful in 
pinpointing a number of important features: 

$\bullet$ Scharf \etal\ (1992) applied spherical harmonics to the 
2-dimensional IRAS PSC
and noted a prominent cluster behind the ZOA in Puppis ($\ell \sim 245\deg$)
which was simultanously discovered as a nearby cluster through 
\HI-observations of obscured galaxies in that region by Kraan-Korteweg
\& Huchtmeier (1992).

$\bullet$ Hoffman (1994) predicted the Vela supercluster at ($280\deg, 
6\deg, 6000$) using 3-dimensional WF reconstructions on the IRAS 1.9 Jy
redshift catalogue (Strauss \etal\ 1992), which was observationally
discovered just a bit earlier by Kraan-Korteweg \& Woudt (1993).

$\bullet$ Using POTENT analysis, Kolatt \etal\ (1995) predicted the center
of the Great Attractor overdensity -- its density peak -- to
lie behind the ZOA at ($320\deg, 0\deg, 4500$). Shortly thereafter,
Kraan-Korteweg \etal\ (1996) unveiled the cluster A3627 as
being very rich and massive and at the correct distance. It hence is
the most likely candidate for the central density peak of the GA.

\subsection{Deeper Reconstructions}
Recent reconstructions have been applied to denser galaxy samples covering
larger volumes (8-10000~\kms) with smoothing scales of the order of
500~\kms\ (compared to 1200~\kms). It therefore seemed of interest
to see whether these reconstructions find evidence for unknown
major galaxy structures at higher redshifts.

The currently most densely-sampled, well-defined galaxy redshift 
catalog is the Optical Redshift Survey (Santiago \etal\ 1995). 
However, this catalog is limited to $|b| \ge 20\deg$ 
and the reconstructions 
(\cf\ Baker \etal\ 1998) within the ZOA are strongly influenced by
1.2~Jy IRAS Redshift Survey data and a mock galaxy distribution 
in the inner ZOA.
We will therefore concentrate on reconstructions based 
on the 1.2~Jy IRAS Redshift Survey only. In the following, 
the structures identified 
in the ZOA by (a) Webster \etal\ (1997) using WF plus
spherical harmonics and linear theory and (b) Bistolas (1998) who
applied a WF plus linear theory and non-constrained realizations on
the 1.2~Jy IRAS Redshift Survey will be discussed and compared 
to observational data.
Fig.~2 in Webster \etal\ displays the reconstructed density fields
on shells of 2000, 4000, 6000 and 8000~\kms; Fig.~5.2
in Bistolas displays the density fields in the ZOA from 1500 to 8000~\kms\
in steps of 500~\kms.

The WLF reconstructions clearly find the recently identified
nearby cluster at ($33\deg, 5\deg-15\deg$, 1500), whereas
Bistolas reveals no clustering in the region of the Local Void out to
4000~\kms. At the same longitudes, the clustering at 7500~\kms\ is 
seen by Bistolas, but not by Webster \etal. The Perseus-Pisces 
chain is strong in both reconstructions, and the 2nd Perseus-Pisces 
arm -- which folds back at $\ell\sim 195\deg$ -- is clearly confirmed.
Both reconstructions find the Perseus-Pisces complex to be very extended 
in space, \ie\ from 3500~\kms\ out to 9000~\kms.
Whereas the GA region is more prominent compared 
to Perseus-Pisces in the Webster \etal\
reconstructions, the signal of the Perseus-Pisces complex 
is considerably stronger than the GA in Bistolas where it does 
not even reveal a well-defined central density peak. Both
reconstructions find no evidence for the suspected PKS1343
cluster but its signal could be hidden in the central (A3627)
density peak due to the smoothing.
While the Cygnus-Lyra complex ($60\deg-90\deg, 0\deg, 4000$) 
discovered by Takata \etal\ (1996) stands out clearly 
in Bistolas, it is not evident in Webster \etal.
Both reconstructions find a strong signal for the Vela SCL 
($285\deg, 6\deg, 6000$), labelled as HYD in WLF. The
Cen-Crux cluster identified by Woudt (1998) is evident in 
Bistolas though less distinct in Webster \etal. A suspected connection at 
($\ell,v) \sim (345\deg, 6000$) -- \cf\ Fig.~2 in Kraan-Korteweg 
\etal\ (1998) --  is supported by both methods.
The Ophiuchus cluster just becomes visible in the most distant
reconstruction shells (8000~\kms).

\subsection{Conclusions}
Not all reconstructions find the same features, and when they do,
the prominence of the density peaks as well as their locations in space
do vary considerable. At velocities of $\sim 4000$~\kms\ 
most of the dominant structures lie close to or within the ZOA while
at larger distances, clusters and voids seem to be more homogeneously
distributed over the whole sky. Out to 8000~\kms\ none of the 
reconstructions predict any major structures which are not 
mapped or suggested from observational data. So no major 
surprises seem to remain hidden in the ZOA. The various 
multi-wavelength explorations of the Milky Way 
will soon be able to verify this. Still, the combination of both
the reconstructed potential fields and the observationally mapped galaxy 
distribution will lead to estimates of the cosmological parameters
$\Omega$ and $b$.

 

\section*{Acknowledgements}

The collaborations with our colleagues in the various
multi-wavelength surveys --- C. Balkowski, V. Cayatte, A.P. Fairall, 
with the redshift follow-ups of the optical surveys, A. Schr\"oder and 
G.A. Mamon in the exploration of the DENIS survey, W.B. Burton, P.A. 
Henning, O. Lahav and A. Rivers
in the northern ZOA \HI-survey (DOGS) and the HIPASS ZOA team members
R.D. Ekers, A.J. Green, R.F. Haynes, P.A. Henning, S. Juraszek,
M. J. Kesteven, B. Koribalski, R.M. Price, E. Sadler and 
L. Staveley-Smith in the southern ZOA survey --- are greatly
appreciated. RKK warmly thanks the University of Western Sydney 
for partial financial suppport.

\section*{References}

\reference
Baker J.E., Davis M., Strauss M.A. \etal\ 1998
ApJ, in press (astro-ph/9802173)

\reference
Bertschinger E., Dekel A., 1989 ApJ 336, 5
\reference
Bistolas V. 1998
Ph.D. thesis, Hebrew University, Jerusalem

\reference
B\"ohringer H., Neumann D.M., Schindler S., Kraan-Korteweg R.C. 1996 
ApJ 467, 168

\reference
Cameron L.M. 1990
A\&A 233, 16

\reference
Chamaraux P., Cayatte V., Balkowski C., Fontanelli P. 1990
A\&A 229, 340

\reference 
Epchtein N. 1997
in The Impact of Large Scale Near-Infrared Surveys, eds.\
F. Garzon \etal, Kluwer, Dordrecht, 15


\reference
Fabian A.C. 1994 
in Unveiling Large-Scale Structures behind the Milky Way, 
eds. C. Balkowski and R.C. Kraan-Korteweg, ASP Conf. Ser. 67, 76
  
\reference
Fisher K.B., Davis M., Strauss M.A. \etal\ 1995
ApJS 100, 69

\reference
Focardi P, Marano B, Vettolani G. 1984 A\&A 136, 178





\reference
Hoffman Y., 1994
in Unveiling Large-Scale Structures behind the Milky Way, 
eds. C. Balkowski and R.C. Kraan-Korteweg, ASP Conf. Ser. 67, 185

\reference
Hudson, M. 1994 
MNRAS 266, 475


\reference
Kolatt T., Dekel A., Lahav O. 1995 
MNRAS 275, 797


\reference
Kraan-Korteweg R.C. 1998
in The Low Surface Brightness Universe, IAU Coll 171, 
eds. J.I. Davies, C. Impey and S. Philipps, A.S.P.Conf. Ser., 
in press (astro-ph/9810255)

\reference
Kraan-Korteweg R.C., Huchtmeier, W.K. 1992
A\&A 266, 150

\reference
Kraan-Korteweg R.C., Woudt, P.A. 1993
in Cosmic Velocity Fields, eds. F. Bouchet and M. Lachi\`eze-Rey, 
Editions Fronti\`eres, 557

\reference
Kraan-Korteweg R.C., Cayatte V., Fairall A.P. \etal\ 1994
in Unveiling Large-Scale Structures behind the Milky Way, 
eds. C. Balkowski and R.C. Kraan-Korteweg, ASP Conf. Ser. 67, 99

\reference
Kraan-Korteweg R.C., Woudt P.A., Cayatte V. \etal\ 1996 
Nature 379, 519


\reference
Kraan-Korteweg R.C., Koribalski B., Juraszek S. 1998
in Looking Deep in the Southern Sky, eds.
R. Morganti, W. Couch, Springer, in press (astro-ph/980466)

\reference
Lahav O. 1994
in Unveiling Large-Scale Structures behind the Milky Way, 
eds. C. Balkowski and R.C. Kraan-Korteweg, ASP Conf. Ser. 67, 171

\reference
Lahav O., Yamada T., Scharf C.A., Kraan-Korteweg, R.C. 1993
MNRAS 262, 711

\reference
Lauberts, A. 1982
The ESO/Uppsala Survey of the ESO (B) Atlas, ESO, Garching

\reference
Lynden-Bell D. 1991 
in Observational Test of Cosmological Inflation, 
eds.~Shanks, T.~\etal, 337 


\reference
Marzke R.O., Huchra J.P., Geller M.J. 1996
AJ 112, 1803

\reference
McAdam W.B. 1991 
PASA 9, 255


\reference
Pantoja C.A., Altschuler D.R., Giovanardi C., Giovanelli R. 1997
AJ 113, 905


\reference
Roman A.T., Takeuchi T.T., Nakanishi K., Saito M. 1998
PASJ 50, 47



\reference
Santiago B.X., Strauss M.A., Lahav O. \etal\ 1995 ApJ 446, 457

\reference
Scharf C., Hoffman Y., Lahav. \etal\ 1992 MNRAS 256, 229

\reference
Schlegel D.J., Finkbeiner D.P., Davis M. 1998 
ApJ 500, 525
 


\reference
Skrutskie, M.F., \etal\ 1997 
in The Impact of Large Scale Near-Infrared Surveys, 
eds.\ F. Garzon \etal, Kluwer, Dordrecht, 25


\reference
Strauss M.A., Huchra J.P., Davis M. \etal\ 1992
ApJS 82, 29

\reference
Strauss M.A., Willick J.A. 1995
Phys. Rep. 26, 27

\reference
Takata T., Yamada T., Saito M. 1996
ApJ 457, 693

\reference
Tashiro M., Kaneda H., Makishima K. \etal\ 1998 
ApJ 499, 713


\reference
Wakamatsu K., Hasegawa T., Karoji H. \etal\ 1994
in Unveiling Large-Scale Structures Behind the Milky Way, 
eds.\ C. Balkowski, R.C. Kraan-Korteweg, ASP Conf. Ser. 67, 131

\reference
Webster M., Lahav O., Fisher K. 1997
MNRAS 287, 425

\reference
West, R.M., Tarenghi M. 1989 
A\&A 223, 61

\reference
Woudt P.A. 1998
Ph.D. thesis, Univ. of Cape Town.






\end{document}